\documentclass[aps,pre,reprint,superscriptaddress]{revtex4-1}
\usepackage{xspace}
\usepackage{graphicx}
\usepackage{footnote}
\usepackage{dcolumn}
\usepackage{amsmath}

\newcommand{\N}{N\'eel\xspace}
\newcommand{\gvs}{$\mathrm{GaV_4S_8}$\xspace}
\newcommand{\Cr}{$\mathrm{Cr_{1/3}NbS_2}$\xspace}
\newcommand{\CuSe}{$\mathrm{Cu_{2}OSeO_3}$\xspace}
\newcommand{\tc}{$T_\mathrm{C}$\xspace}
\newcommand{\tsub}[1]{$T_{\mathrm{#1}}$\xspace}
\newcommand{\rem}[1]{$M_{#1\omega}^{\prime}$\xspace}
\newcommand{\imm}[1]{$M_{#1\omega}^{\prime\prime}$\xspace}
\newcommand{\cpxm}[1]{$M_{#1\omega}$\xspace}
\newcommand{\resuscept}[1]{$\chi_{#1\omega}^{\prime}$\xspace}
\newcommand{\imsuscept}[1]{$\chi_{#1\omega}^{\prime\prime}$\xspace}
\newcommand{\suscept}[1]{$\chi_{#1\omega}$\xspace}

\begin{document}

\title{Robust cycloid crossover driven by anisotropy in the skyrmion host GaV$_\mathbf{4}$S$_\mathbf{8}$}

\author{E. M. Clements}
\email[Corresponding author: ]{emclements@mail.usf.edu}
\affiliation{Department of Physics, University of South Florida, Tampa, Florida 33620, USA}

\author{R. Das}
\affiliation{Faculty of Materials Science and Engineering and Phenikaa Institute for Advanced Study (PIAS), Phenikaa University, Hanoi 10000, Vietnam}
\affiliation{Phenikaa Research and Technology Institute (PRATI), A\&A Green Phoenix Group, 167 Hoang Ngan, Hanoi 10000, Vietnam}

\author{G. Pokharel}
\affiliation{Department of Physics and Astronomy, University of Tennessee, Knoxville, Tennessee 37996, USA}
\affiliation{Materials Science \& Technology Division, Oak Ridge National Laboratory, Oak Ridge, Tennessee 37831, USA}

\author{M. H. Phan}
\email[Corresponding author: ]{phanm@usf.edu}
\affiliation{Department of Physics, University of South Florida, Tampa, Florida 33620, USA}

\author{A. D. Christianson}
\affiliation{Materials Science \& Technology Division, Oak Ridge National Laboratory, Oak Ridge, Tennessee 37831, USA}

\author{D. Mandrus}
\affiliation{Department of Physics and Astronomy, University of Tennessee, Knoxville, Tennessee 37996, USA}
\affiliation{Materials Science \& Technology Division, Oak Ridge National Laboratory, Oak Ridge, Tennessee 37831, USA}
\affiliation{Department of Materials Science and Engineering, University of Tennessee, Knoxville, Tennessee 37996, USA}

\author{J. C. Prestigiacomo}
\affiliation{U.S. Naval Research Laboratory, Washington D.C., 20375, USA}
\author{M. S. Osofsky}
\affiliation{U.S. Naval Research Laboratory, Washington D.C., 20375, USA}

\author{H. Srikanth}
\email[Corresponding author: ]{sharihar@usf.edu}
\affiliation{Department of Physics, University of South Florida, Tampa, Florida 33620, USA}

\date{\today}

\begin{abstract}
We report on the anomalous magnetization dynamics of the cycloidally-modulated spin textures under the influence of uniaxial anisotropy in multiferroic \gvs. The temperature and field dependence of the linear ac susceptibility [\resuscept1{}($T,H$)], ac magnetic loss [\imsuscept1{}($T,H$)], and nonlinear ac magnetic response [\cpxm3{}($T,H$)] are examined across the magnetic phase diagram in the frequency range $f = 10-10000$ Hz. According to recent theory, skyrmion vortices under axial crystal symmetry are confined along specific orientations, resulting in enhanced robustness against oblique magnetic fields and altered spin dynamics. We characterize the magnetic response of each spin texture and find that the dynamic rigidity of the \N skyrmion lattice appears enhanced compared to Bloch-type skyrmions in cubic systems, even in the multidomain state. Anomalous \cpxm3 and strong dissipation emerge over the same phase regime where strong variations in the cycloid pitch were observed on lowering temperature in recent small-angle neutron scattering experiments  [White et al., Phys. Rev. B 97, 020401(R) (2018)]. Here, we show that strong anisotropy also drives an extended crossover of the zero-field cycloid texture in \gvs. The frequency dependence of these dynamic signatures is consistent with that of a robust anharmonic spin texture exhibiting a correlated domain arrangement. The results underpin the essential role of magnetic anisotropy in enhancing the rigidity of topological spin textures for diverse applications.
\end{abstract}

\maketitle

\section{Introduction \label{sec:intro}}
\begin{figure*}[]
\includegraphics[width=\linewidth]{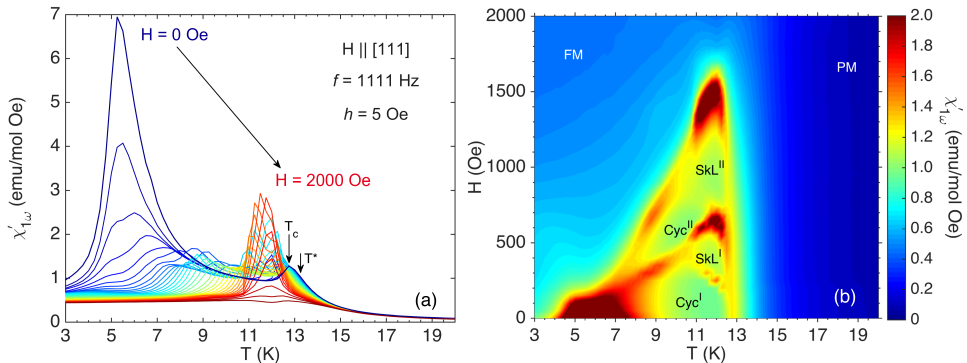}
\caption{Temperature and field dependence of the ac susceptibility, \resuscept1{}$(T,H)$.}
\label{fig:linchi}
\end{figure*}

	Topologically nontrivial spin textures have been the focus of intense study since the experimental observation of the magnetic skyrmion---a magnetic soliton with a vortex-like spin arrangement---in a chiral magnet \cite{muh09}. Particularly, the nanometer size and localized nature of topological magnetic solitons, such as the 1D kink and 2D skyrmion, hold promise for applications in spintronics \cite{kis15,fer17,gar17,man04}, information storage \cite{kis11}, and computing \cite{pin18,sin19}. Solitonic spin textures may be stabilized by a variety of mechanisms brought on by competing interactions \cite{lin16,leo15,bat16}, often in the presence of thermal fluctuations \cite{muh09}, and may be further enhanced by magnetic anisotropic effects \cite{leo17,lin16}.

	In Dzyaloshinskii-Moriya interaction (DMI)-stabilized spin textures, the antisymmetric exchange interaction imparts stability to the phase, $\phi(r)$, of the spatially-varying order parameter, e.g. $M\propto \exp[\pm i\phi(r)]$ \cite{tog16}, where the sign defines the handedness of the spin rotation. The fixed rotation sense \cite{bog94} leads to robust spin textures with periodicity that extends over large length scales, even across material defects \cite{tog12}.  As a result, the magnetization processes in these materials require rearrangements of essentially macroscopic spin textures, leading to very slow dynamics \cite{bau16a,bau16b,bau17}, extended transitions, and large magnetic losses \cite{lev14,cle18}. Thus, the analysis of ac magnetization has been employed extensively in the chiral helimagnets (CHM) as a powerful tool to capture details of the phase evolution due to the unique dynamic properties of long-wavelength magnetic structures. \cite{bau12,lev14,qia16,cle18}

	Another anomalous dynamic signature arises when highly coherent spin textures are subjected to a harmonic magnetic field and the periodic magnetic response curve, $M(t)$, exhibits strong anharmonicity. Due to its relation to spatial symmetry breaking, the leading odd harmonic of the ac magnetic response (\cpxm3), has been used phenomenologically to probe the character of phase transitions, namely ferromagnetic (FM), spin glass, and antiferromagnetic transitions \cite{suz77,fuj81,bal13}. At very low frequencies, $\alt 10$ Hz, \cpxm3 reflects the dynamics of magnetic domains, as pointed out and thoroughly explored by Mito and coworkers \cite{mit09,mit12,mit15,tsu16b,tsu18}. Motivated by the description of magnetization processes for small amplitude fields by Rayleigh \cite{ray87}, where the domain wall displacement under a harmonic force could be described by in-phase, out-of-phase, and odd harmonic components, they established a diagnostic approach that characterizes domain dynamics into five types \cite{mit15}. Here, the latter two components reflect the irreversible Barkhausen jumps as domain walls overcome energy barriers in an ac process. 
	
	By expanding the usual treatment of a magnetic domain wall in terms of a driven damped harmonic oscillator \cite{bal03}, the anharmonic spring model was applied to account for nonlinear contributions to the response \cite{mit15}: 
	\begin{equation}\label{eqn:Duffing}
	\frac{d^2x}{dt^2} + 2\gamma \frac{dx}{dt} + \omega^2_0x + \eta x^3 = F \sin (\omega t),
	\end{equation}
where $\gamma$ is the damping parameter, $\omega_0 = (k/m)^{1/2}$ is the (undamped) natural frequency for a domain wall with effective mass, $m$, attached to a spring with spring constant, $k$, and $\eta$ controls the nonlinear term representing variations in spring stiffness depending on the amplitude of displacement. Replacing $x$ with the magnetic response (i.e. the deviation from the equilibrium magnetic state) and the righthand side with an oscillating magnetic field, Eq.{} \ref{eqn:Duffing} describes the shape of an ac hysteresis loop. On the lefthand side, the first term is connected to the domain wall inertia, the damping term to any magnetic dissipation, and the third term to stiffness arising from anisotropy and pinning effects. The fourth term, $\eta x^3$, crucially accounts for the \cpxm3 component of the ac magnetic response. 

	The phenomenological model shown above was first developed to diagnose domain dynamics from ac hysteresis loops of itinerant helical magnet MnP \cite{mit15}, which exhibited anomalously large anharmonicity in various regions of the magnetic phase diagram. Analyses of both the damping and nonlinear contributions to the total magnetization response---mainly via the magnetic dissipation, \imm1, and \cpxm3, respectively---has since been applied to a variety of materials with modulated spin structures \cite{mit09,mit12,mit15,tsu16b,tsu18,cle18}. In the case of a crystalline material lacking an inversion center, the DMI induces anisotropy over the entire crystal. As a result, the enhanced thermodynamic rigidity of the structure \cite{tog16} leads to large amplitude contributions from \cpxm3 and the magnetic loss, the character of which depends on the spin configuration.

As distinct features of $M(t)$ serve as a type of dynamic finger-print of the magnetic structure, the following study characterizes the incommensurate spin textures in the multiferroic lacunar spinel \gvs under the influence of uniaxial magnetocrystalline anisotropy. A cooperative Jahn-Teller distortion at \tsub{JT}$= 42$ K stretches its tetrahedral V$_\mathrm{4}$ clusters (each carrying spin = 1/2) along any of four possible $\left<111\right>$ directions, reducing the symmetry from cubic $F\bar43m$ to polar $R3m$ \cite{poc00, but17b}.  Below the Curie temperature, \tc $= 13$ K, the long-range magnetic order is controlled by competition between DMI, ferromagnetic exchange, easy-axis anisotropy along the direction of ferroelectric polarization, and the Zeeman energy \cite{kez15}. \gvs differs from the archetypal cubic skyrmion lattice (SkL) hosts in an important way: its axially symmetric $C_{3v}$ crystal structure confines the \N-type skyrmion cores to the rhombohedral easy axis \cite{kez15, ehl16}. The orientation of the \N SkL, which can be described by a superposition of cycloidal spin modulations that propagate radially from the core \cite{whi18}, is rigid with respect to tilted magnetic fields, unlike skyrmions in cubic system which orient along the field direction \cite{leo17}.  Additionally, the incommensurate to commensurate (IC--C) transitions are controlled by the strength of the uniaxial anisotropy that becomes the dominant energy scale at low temperatures.

So far, the analysis of the nonlinear magnetic response in systems hosting magnetic solitons has been restricted to materials with chiral crystal structures \cite{mit09,mit12,mit15,tsu16b,tsu18,cle18}. While the chiral spin interactions induced by the DMI lead to a fixed rotation sense, the cycloidally-modulated states in polar \gvs do not exhibit macroscopic chirality. Thus, \gvs offers a unique opportunity to study domain dynamics of achiral spin phases that are topologically nontrivial. In the following, we investigate the nonlinear ac magnetic response and magnetic dissipation phenomena across the phase boundaries of the noncollinear magnetic states in \gvs from the paramagnetic (PM) phase down to $T = 3$ K for magnetic fields to $H = 2000$ Oe. The magnetization dynamics are characterized in a frequency range $f = 10 - 10000$ Hz for the magnetic cycloid (Cyc)--PM, Cyc--SkL, SkL--FM, Cyc--FM transitions with special emphasis on the anisotropy-driven IC--C transition at low temperatures.

\begin{figure*}[]
\includegraphics[width=\linewidth]{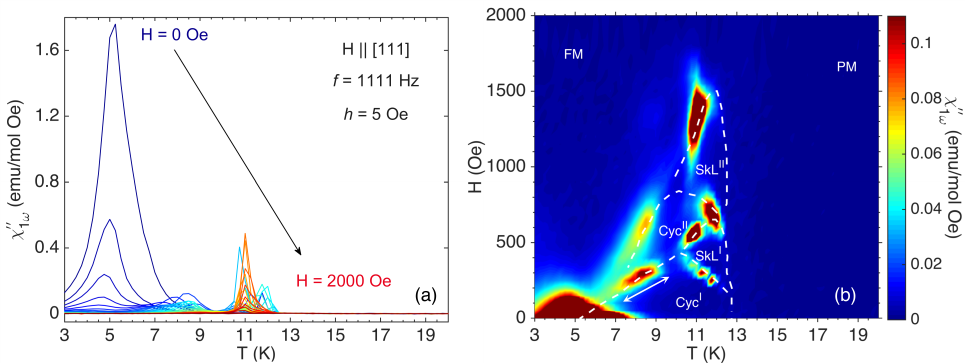}
\caption{Temperature and field dependence of the ac magnetic loss, \imsuscept1{}$(T,H)$.}
\label{fig:loss}
\end{figure*}
\section{Experimental Methods \label{sec:exp}}

Single crystals of \gvs were grown by a chemical vapor transport method using I$_2$ gas similar to that described in Ref. \cite{war17}. Static and dynamic magnetization measurements were performed using a commercial Physical Property Measurement System (PPMS, Quantum Design) with a vibrating sample magnetometer (VSM) and an ac measurement option (ACMS), respectively. The single crystals were oriented such that magnetic field, $H$, was applied along principle cubic directions $[111], [110]$, and $[100]$. 

For the dc measurements, thermal hysteresis was measured using zero-field cooled, field-cooled-warming, and field-cooled-cooling (FCC) protocols for selected magnetic fields. Unless otherwise noted, ac measurements were collected as a function of temperature for $T = 3 - 20$ K in 0.25 K steps in fixed dc fields, $H = 0 - 2000$ Oe, in 50 Oe steps under FCC conditions. Between each temperature sweep, the sample was warmed to $T = 50$ K, well above the ferromagnetic Curie temperature, \tc $= 13$ K, and above the temperature of the ferroelectric Jahn-Teller distortion, \tsub{JT} $= 42$ K.

The ac driving field, $h_\mathrm{ac}=h \sin{\omega t}$, where $\omega = 2\pi f$, was varied in the frequency range $f = 10 - 10000$ Hz with amplitude $h = 5$ Oe. The time-dependent magnetic response to $h_\mathrm{ac}(t)$ can be expanded as \cite{bal03}
\begin{eqnarray} 
&M (t) =M_{1\omega} \sin (\omega t + \theta_{1\omega}) + M_{2\omega} \sin (2\omega t + \theta_{2\omega}) \nonumber \\ 
&+M_{3\omega} \sin(3\omega t + \theta_{3\omega}) + ...,
\label{eqn:harmonics}
\end{eqnarray}
where $M_{n\omega}$ is the $n^{th}$ harmonic component (for integer $n$ = 1, 2, 3, ...) of the magnetic response and $\theta_{n\omega}$ is the delay in phase of each component against $h_\mathrm{ac}$. $M_{n\omega}$ components were recorded using a lock-in technique. The in- and out-of-phase components of the linear ac susceptibility, \suscept1 = \cpxm1/h, are given by \resuscept1 = \suscept1$\cos\theta_{1\omega}$ and \imsuscept1 = \suscept1$\sin\theta_{1\omega}$, respectively. 

\section{Results \label{sec:results}}

\subsection{Magnetic phase diagram \label{sec:magphasedia}}

Figure \ref{fig:linchi}(a) presents the ac susceptibility measured as a function of temperature in fixed dc magnetic fields, $\chi^{\prime}_{1\omega}(T,H)$, at $f = 1111$ Hz for $H$ applied along the pseudocubic $[111]$ direction, the magnetic easy axis of one of the four structural variants of the crystal. Due to small changes in magnetization across phase transformations between spiral, SkL, and homogenous spin states, maxima in \resuscept1 are typically utilized to mark the phase boundaries \cite{bau12,lev14,qia16,but17a}. In particular, the discontinuous formation and annihilation of magnetic vortices, first described theoretically by Bogdanov \cite{bog89}, physically appears in real systems as a finite peak in \resuscept1 on either side of the SkL phase pocket. The $H-T$ intensity map of \resuscept1 is shown in Fig.~\ref{fig:linchi}(b). Indeed, the maxima in \resuscept1 delineate the boundaries between modulated magnetic cycloid and SkL phase pockets, in agreement with Butykai, et al. \cite{but17a}. Due to easy axis anisotropy, the effective magnetic field in each domain scales by the direction cosine of the applied field with respect to the easy axis. Thus, two additional phases pockets are observed at higher fields in Fig.~\ref{fig:linchi}(b), labelled Cyc$^\mathsf{II}$ and SkL$^\mathsf{II}$, where structural domains stretched along [\=111],[1\=11],[11\=1] all span $71^\circ$ with $H$. Moving toward low temperature, $\chi^{\prime}_{1\omega}(T,H)$ tracks the anisotropy-driven transitions from the incommensurate spin textures to the commensurate ferromagnetic state. The IC--C transitions are centered around critical temperatures $T_\mathrm{IC\rightarrow C}(H)$ that increase with $H$ from $T=5$ K at $H=0$ to $T=11.75$ K at $H = 1500$ Oe. Above $H = 1700$ Oe, a relatively shallow hump near \tc $=13$ K shifts to higher temperature with $H>2000$ Oe, separating the FM and PM phases.

	The broadened anomalies in the $H-T$ phase diagram signify complex and extended phase transitions across the IC--C phase boundaries at $T_\mathrm{IC\rightarrow C}(H)$, particularly for $H < 200$ Oe. First, we consider the zero-field temperature evolution of the in-phase susceptibility shown in Fig.~\ref{fig:linchi}(a). As temperature is lowered from 20 K in the paramagnetic state \resuscept1 rises, reaches a point of inflection at $T^*=13.25$ K, and finally a kink point at \tc = 12.75 K. Similar to the chiral helimagnets \cite{sti08}, the kink point identifies the onset of spiral long-range order (LRO). In polar \gvs, the LRO takes the form of a magnetic cycloid which propagates within the $\{111\}$ planes, along the set of directions $\left<110\right>$ \cite{whi18}.  After decreasing in the temperature range $12.75 \: \mathrm{K} \ge T \ge 12$ K, \resuscept1 rises to another peak at $T_\mathrm{IC\rightarrow C}(0) = 5.25$ K. The broad anomaly reaches the susceptibility value of the kink point ($1.28$ emu/mol Oe) near $T = 9$ K and again near $T = 3.75$ K, spanning a large temperature range relative to $T_\mathrm{IC\rightarrow C}(0)$. 

	In the literature, the peak center near $T \approx 5$ K is often identified as the transition into ferromagnetic order \cite{ruf15,wid17}. Recent small-angle neutron scattering (SANS) results by White, et al. provide microscopic evidence that the cycloid pitch gradually stretches as temperature is lowered below \tc \cite{whi18}. The IC--C process likely terminates at a ferromagnetic ground state as easy axis anisotropy becomes the dominant energy scale at low temperature. In accordance with the SANS results, the transition centered at $T_\mathrm{IC\rightarrow C}(0)$ shown in Figs.~\ref{fig:linchi}(a--b) is extended in nature. However, as demonstrated in later sections, the transition may not be complete down to $T = 3$ K. 

\subsection{Dissipation mechanisms \label{sec:dissmech}}

\begin{figure}[t]
\includegraphics[width=\linewidth]{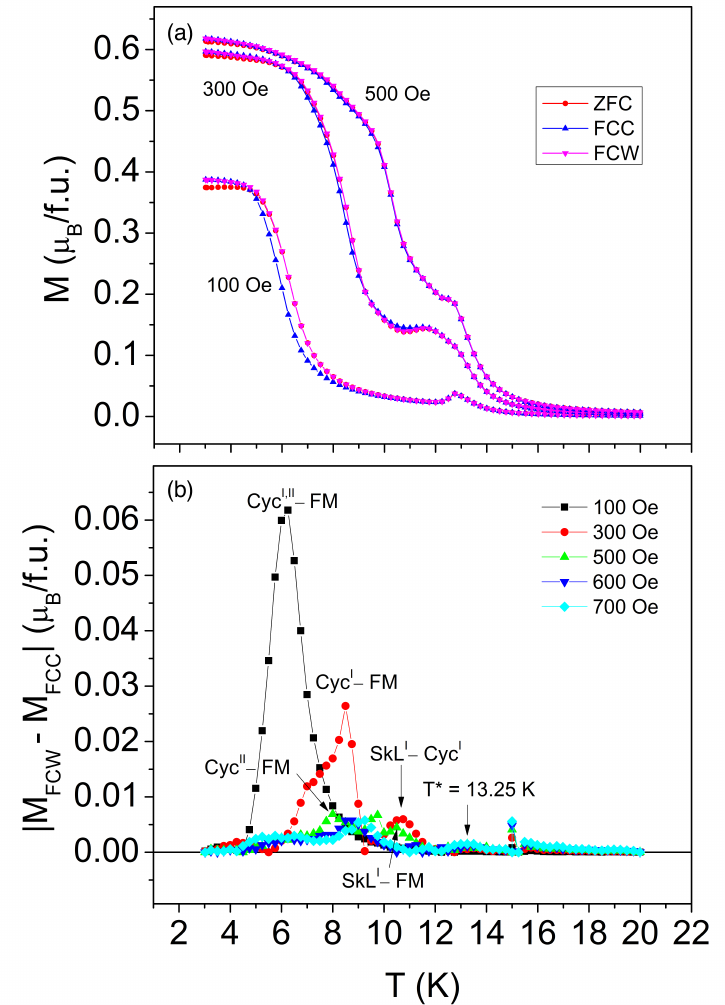}
\caption{Thermal hysteresis of the static magnetization (a) under zero-field cooled, field-cooled cooling, and field-cooled warming protocols for $H = 100, 300, 500$ Oe. (b) Difference between the magnetization between cooling and warming curves.}
\label{fig:thermhyst}
\end{figure}

	The out-of-phase component of the ac susceptibility, \imsuscept1{}$(T,H)$, is shown in Fig.~\ref{fig:loss}(a) and its associated intensity plot is shown in Fig.~\ref{fig:loss}(b). Dashed lines track the magnetic phase boundaries, as defined above from maxima in \resuscept1{}$(T,H)$, and demonstrate that the broad maxima of the magnetic dissipation profile tend to shift toward lower temperatures compared to the in-phase component. Despite the stark contrast between the phase diagrams of the chiral helimagnets and polar \gvs, the loss profile bears some similarities to other SkL hosts. Most notably, \imsuscept1 is minimal on entering the modulated states from high temperature \cite{lev14} and at the vertex of the low temperature SkL boundary \cite{tsu18}, which in \gvs coincides with the SkL--Cyc--FM triple point (10.25 K, 450 Oe). At the boundaries of the PM phase, our instrument resolves an out-of-phase ac moment on the order of $10^{-7}$ emu along the PM--SkL border, above the instrument resolution of $\sim10^{-8}$ emu. Furthermore, magnetic loss is virtually absent in the pure phases, but accompanies the transitions at the upper and lower field boundaries of the SkL pocket, namely between the modulated states (SkL$^\mathsf{I,II}$--Cyc$^\mathsf{I,II}$) and into the FM state (SkL$^\mathsf{I,II}$--FM), respectively. Similar behavior is observed on lower and upper phase boundaries of the Bloch-type SkL pocket in \CuSe, which is embedded in a conical phase \cite{lev14}.

	Anomalous magnetic loss appears across the zero-field Cyc--FM transition [Figure \ref{fig:loss}(a)] where, presumably, the spin cycloid in all four domains undergo the anisotropy-driven IC--C transition \cite{ruf15,kez15,wid17,whi18}. Here, however, the maximum at $T_\mathrm{IC\rightarrow C}(0) = 5.25$ K drops rapidly with applied magnetic field and disappears near $H=200$ Oe. To analyze the possible dissipation mechanisms that contribute to \imsuscept1, we begin with thermal hysteresis measurements, which may point to (a) metastability in mixed-phase regimes or (b) slow time-scales of the phase transformations. Figure \ref{fig:thermhyst}(a) displays the static magnetization as a function of temperature and field, $M(T,H)$, recorded under warming and cooling protocols, as described in Sec.~\ref{sec:exp}. A plot of $|M_\mathrm{FCW}-M_\mathrm{FCC}|$ vs $T$ in Fig.~\ref{fig:thermhyst}(b) quantifies the hysteresis in warming and cooling curves for selected magnetic fields. Cyc--FM transitions display strong bifurcation especially for $H\le300$ Oe. While the hysteretic contributions from the SkL--FM and SkL--Cyc transitions are much smaller, $|M_\mathrm{FCW}-M_\mathrm{FCC}|$ vs $T$ remains nontrivial, notably across the SkL$^\mathsf{I}$--Cyc$^\mathsf{I}$ phase boundary at $H = 300$ Oe. Additionally, thermal hysteresis develops with increasing magnetic field at $T^* = 13.25$ K, which represents the inflection point in $M(T)$. 

\begin{figure}[b]
\includegraphics[width=\linewidth]{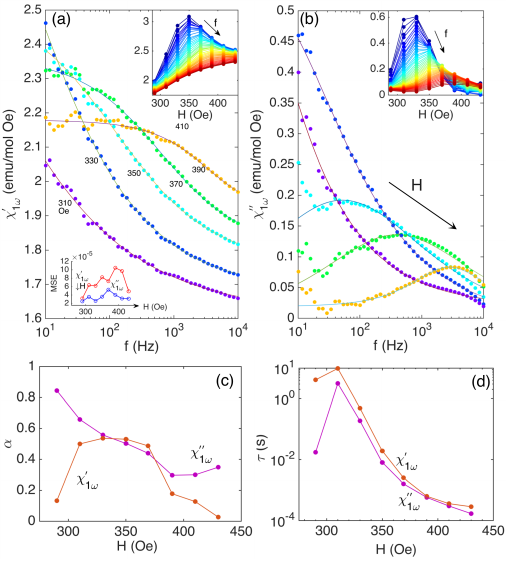}
\caption{The frequency dependence of (a) \resuscept1 and (b) \imsuscept1 at $T = 10.75$ K for magnetic fields spanning the Cyc$^\mathsf{I}$--SkL$^\mathsf{I}$ mixed-phase regime across the phase boundary fit to Eqns.~(\ref{eqn:realcole}) and (\ref{eqn:imagcole}), respectively. Top insets in (a) and (b) display the peak shifts to higher field with increasing frequency. Lower inset in (a) display the mean squared error for fits at each $H$. (c) Parameter, $\alpha$, characterizing the symmetric distribution of relaxation times about the (d) average time, $\tau_0$, from fits of (a--b) to the Cole-Cole model.}
\label{fig:Cole}
\end{figure}

\begin{figure}[]
\includegraphics[width=\linewidth]{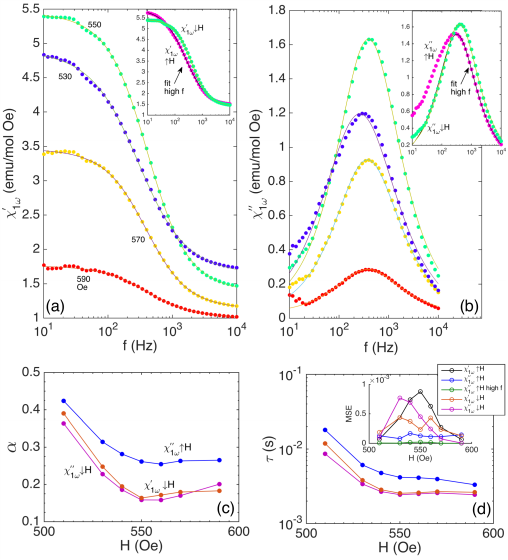}
\caption{The frequency dependence of (a) \resuscept1 and (b) \imsuscept1 at $T = 10.75$ K for descending magnetic fields ($H_\downarrow$) spanning the SkL$^\mathsf{I}$--FM mixed-phase regime across the phase boundary fit to Eqns.~(\ref{eqn:realcole}) and (\ref{eqn:imagcole}), respectively. Insets in (a) and (b) compare curves from ascending ($H_\uparrow$) and descending field protocols.  (c) Parameter, $\alpha$, characterizing the symmetric distribution of relaxation times about the (d) average time, $\tau_0$, from fits of (a--b) to the Cole-Cole model. The blue curves show dynamic parameters extracted from the fit to the entire frequency range in the ascending measurement.}
\label{fig:skydyn}
\end{figure}

	Thermal hysteresis may signal the discontinuous nature of skyrmion formation, as predicted by Bogdanov et al. for magnetic field-induced vortex formation \cite{bog89}. Indeed, previous ac susceptibility studies attribute glassy behavior to mixed magnetic phases  \cite{lev14,qia16,but17a} where, in light of Lorentz and magnetic force microscopy studies, the phenomena were related to the nucleation of short-range skyrmions as topological defects within the longer-ranged FM phase \cite{uch06,mil13}. As shown previously by Butykai et al. in \cite{but17a}, the transitions in \gvs display a broad distribution of relaxation times. Figs.~\ref{fig:Cole}(a--b) and \ref{fig:skydyn}(a--b) display \resuscept1$(f,H)$ and \imsuscept1$(f,H)$ for Cyc--SkL and SkL--FM transitions, respectively, at $T = 10.75$ K with fits to the Cole-Cole model as follows. The measurements were performed as a function of ascending field ($H_\uparrow$) after zero-field cooling and descending field ($H_\downarrow$) after reaching $H = 2000$ Oe. The dynamics are tracked along the susceptibility peak, to observe changes on either side of the transitions, namely moving from the Cyc into the SkL and from the SkL to the FM phase.
	
The Cole-Cole modification \cite{col41} of the Debye model,
\begin{equation}
 \chi(\omega) = \chi_{\infty} + \frac{\chi_0 - \chi_{\infty}}{1 + (i \omega \tau)^{1-\alpha}}, 
\label{eq:cole}
\end{equation}
introduces the parameter $\alpha$ to account for a distribution of relaxation times: $\alpha = 1$ corresponds to an infinitely broad distribution and $\alpha = 0$ accounts for a single relaxation process. $\chi_0$ is the susceptibility in the limit of low frequency ($\omega \rightarrow 0$), where heat exchange in the magnetic system occurs between the spins and lattice vibrations in an isothermal process. $\chi_{\infty}$ is the susceptibility in the high frequency limit ($\omega \rightarrow \infty$), where the spins remain isolated from their surroundings and relaxation occurs via an adiabatic process \cite{mor01,bal03}. The Cole-Cole model assumes a logarithmic distribution of relaxation times, $\tau$, which is symmetric about the characteristic or average $\tau_0 = 1/(2 \pi f_0)$. $\chi(\omega)$ can be decomposed into in- and out-of-phase components,
\begin{equation}
\chi^{'}(\omega) = \chi_{\infty} +\frac{[\chi_{0} - \chi_{\infty}] [1+(\omega \tau_0)^{1-\alpha} \sin(\pi \alpha/2)]}{1+2(\omega\tau_0)^{1-\alpha} \sin(\pi \alpha /2) + (\omega \tau_0)^{2(1-\alpha)}},
\label{eqn:realcole}
\end{equation}
\begin{equation} 
 \chi^{''}(\omega) =   \frac{[\chi_0 - \chi_{\infty}] (\omega \tau_0)^{1-\alpha} \cos(\pi \alpha/2)}{1+2(\omega\tau_0)^{1-\alpha} \sin(\pi \alpha /2) + (\omega \tau_0)^{2(1-\alpha)}}. 
\label{eqn:imagcole}
\end{equation}
	For the Cyc--SkL transition, fits to the in- and out-of-phase susceptibility for $H_\downarrow$ yield slightly different parameter sets, especially at lower and upper field boundaries. No significant differences were observed between data in ascending and descending fields. The transition takes on a glass-like character, according to field dependence of $\alpha$, which reaches values of up to 0.8, indicating a stretched exponential relation. The average relaxation time appears to drop with $H$, moving from the low-field side (Cyc-dominated) into the high-field side (SkL-dominated) of the transition peak, eventually falling outside of our time window ($\sim10^{-4}$ s). For $H\ge350$ Oe, a hump in \imsuscept1 forms near 50 Hz while a low-frequency tail remains, suggesting an additional dynamic process exists at much longer timescales. Indeed, Butykai et al., observed an average relaxation time on the order of $10^0$ s at $T = 10.75$ K across the Cyc--SkL transition. These coexisting processes, separated in timescale, agree with the presence of mixed phase behavior expected at a first-order transition.
	
	On the other hand, for the SkL--FM transition, evidence of coexisting processes takes on a different form in \imsuscept1$(f,H)$. Instead of two peaks separated in frequency, on the ascending measurement \imsuscept1 is asymmetric about the peak ($\tau_0$), as shown in the inset of Fig.~\ref{fig:skydyn}(b), where the black line represents the fit to the high frequency side of \imsuscept1{}$(f)$. After increasing the field to $H = 2000$ Oe, the descending measurement displays the symmetry expected in the Cole-Cole formalism (gold fit line). Hysteretic behavior appears as a shift in relaxation time to faster timescales in the descending process as well as a reduction in magnitude of the loss, reflecting differences between annihilation and nucleation processes across the SkL--FM transition. A low frequency tail of enhanced susceptibility has also been observed in the conical-SkL mixed phase in $\mathrm{Fe_{1-x}Co_xSi}$ (x = 0.30) \cite{ban16} and in \Cr, where a lattice of 1D chiral magnetic solitons separate FM domains \cite{cle18}. In both cases the behavior was attributed to an asymmetric distribution of relaxation times. More generalized models taking into account the asymmetry of \imsuscept1{}$(f)$ on a logarithmic scale, such as the Cole-Davidson or Havriliak-Negrami equations, also fail to describe the frequency dependence, as explored in Ref.~\cite{but17a} for the Cyc--FM transition in \gvs. 
	
	 In addition to the asymmetric loss profiles mentioned above, discrepancies in the parameter sets extracted from fits of \resuscept1 and \imsuscept1 to Eqs.~\ref{eqn:realcole} and \ref{eqn:imagcole}, respectively, [as in Fig.~\ref{fig:Cole}(c)] have also been reported by Qian, et al. for CHM-conical and conical-SkL transitions in \CuSe \cite{qia16}.  While the anomalous departures from the Cole-Cole model vary between material systems, the phenomena have all been linked to coexisting dynamic processes from small and large lengthscales. However, a full theoretical treatment accounting for the collective dynamics of the macroscopic spin texture remains unexplored in these systems.

\begin{figure}[b]
\includegraphics[width=\linewidth]{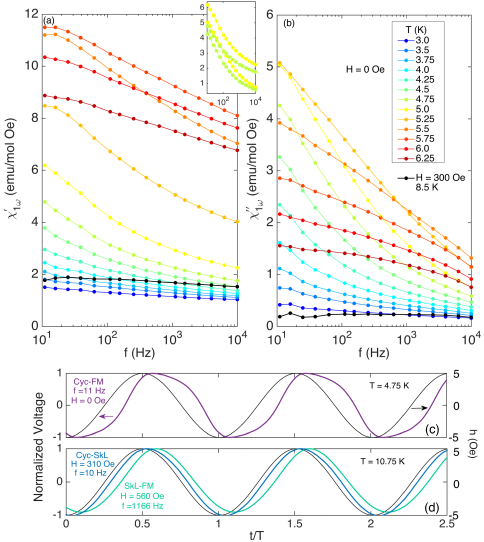}
\caption{Frequency dependence of (a) \resuscept1 and (b) \imsuscept1 for temperatures spanning the extended incommensurate Cyc to commensurate FM transition at zero-field. For comparison, \resuscept1{}$(f)$ and \imsuscept1{}$(f)$ across the Cyc$^\mathsf{I}$--FM transition ($H=300$ Oe, $T=8.5$ K) displays weak frequency dependence. The inset in (a) compares \resuscept1{}$(f)$ and \imsuscept1{}$(f)$ for $T=4.75$ K and $T=5$ K. (c) The normalized response voltage to $h_{ac} = h \sin \omega t$, where $h = 5$ Oe, across the zero-field Cyc--FM transition is highly distorted while (d) Cyc--SkL and SkL--FM transitions show predominantly linear behavior.}
\label{fig:Cycdynam}
\end{figure}
\begin{figure*}[]
\includegraphics[width=\linewidth]{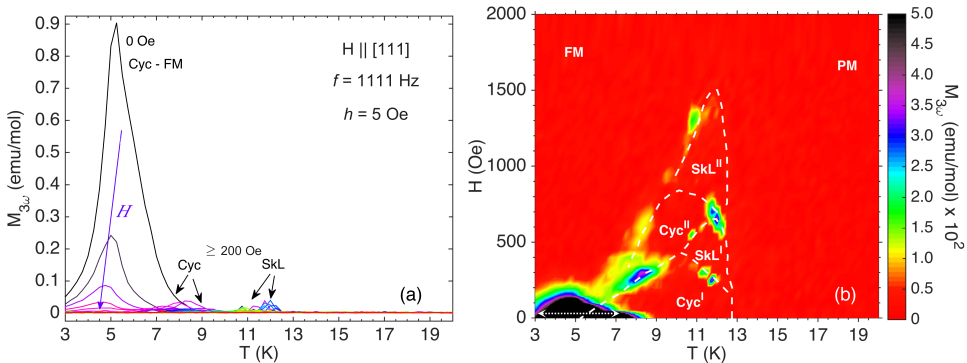}
\caption{Temperature and field dependence of the third harmonic magnetic response, \cpxm3{}$(T,H)$.}
\label{fig:M3}
\end{figure*}
	Ultra-slow rearrangements of the spin structures, such as motion of coherent domain structures, may also contribute to the observed irreversibility in \gvs, especially at the boundaries between the FM phase as shown for the Cyc--FM transition in Fig.~\ref{fig:thermhyst} and the SkL--FM transition in Fig.~\ref{fig:skydyn}. Reports of slow magnetization processes are ubiquitous in the literature for Dzyaloshinskii-Moriya interaction-stabilized spin textures. The long-range order of both the amplitude and phase of the spatially-varying order parameter underlies the observed phenomena; The magnetic structures are coherent over large length scales and the magnetization dynamics reflects the response of a macroscopic texture. Indeed, in Ref.~\cite{but17a}, Butykai et al. showed that across the SkL$^\mathsf{I}$--FM and SkL$^\mathsf{I}$--Cyc$^\mathsf{I}$ mixed-phase regimes in \gvs, $\tau_0$ dropped dramatically with only a small temperature step, reaching the minutes scale by $T =  10$ K and virtually freezing below the triple point at $T = 9.5$ K.  

	As illustrated in Figs.~\ref{fig:Cycdynam}(a--b), below the triple point the magnetic loss stretching along the Cyc$^\mathsf{I,II}$--FM boundaries for $H>200$ Oe and $T=7 - 9.25$ K [double arrows in Fig.~\ref{fig:loss}(b)] shows virtually no frequency dependence. The ultra-slow dynamic state exhibits a flat-lined \resuscept1 and \imsuscept1 vs $f$ (black curve), shown here at $T = 8.5$ K for $H = 300$ Oe. However, the behavior of the zero-field IC--C transition displays strong frequency dependence in both real and imaginary parts (colored curves), spanning temperatures in the range 7 K $\le T \le$ 3 K---at much lower temperatures than the Cyc$^\mathsf{I,II}$--FM transitions at elevated fields. In zero field the \resuscept1 and \imsuscept1 vs $f$ curves do not follow the conventional Cole-Cole-like behavior, in accordance with behavior observed for Cyc--FM transitions slightly below the triple point in Ref.~\cite{but17a}. For example, near $T_\mathrm{IC\rightarrow C}(0)$ a sigmoid-like frequency dependence shows up for both real and imaginary parts, which are almost equal in magnitude (inset). Significant differences between the dissipation profiles of Cyc--FM for $T_\mathrm{IC\rightarrow C}(H<200$ Oe) and $T_\mathrm{IC\rightarrow C}(H>200$ Oe) are also evident across the phase diagram in Figure \ref{fig:loss}(b). At elevated fields, the loss lies along the phase boundaries (as shown by double arrows), however, for $H = 0$, the loss extends deep into the Cyc phase and stretches from around $T=8$ K down to at least $T=3$ K. Taken together, the disparities between the magnetization dynamics for the Cyc--FM transitions at low and elevated fields point to separate origins for their behavior. 
	
	Figures \ref{fig:Cycdynam}(c--d) compare the normalized response voltage to the ac magnetic field, $H_\mathrm{ac}(t)$, for the zero-field Cyc--FM transition and the Cyc--SkL and SkL--FM transition regimes evaluated in Fig.~\ref{fig:Cole}. As expected from the nonzero \imsuscept1 across these transitions presented above, all three responses show delays in phase ($\theta_{1\omega}$) against the ac field, $h\sin \omega t$. However, significant differences exist between the line profiles of the responses in (c) and (d). By inspection, the periodic curves in (d) appear to be dominated by a linear (harmonic) response to the sinusoidal driving field. In contrast, near $T_\mathrm{IC\rightarrow C}(0)$, the periodic curve becomes strongly distorted [Fig.~\ref{fig:Cycdynam} (c)]. The implications of the anharmonic behavior seen here, coupled with the dissipative behavior presented above, will be explored in the following section by looking closely at the leading nonlinear component of the total ac magnetic response in different regimes of the phase diagram. 

\subsection{Nonlinear magnetic response \label{sec:nonlin}}

	In a study of the itinerant helical magnet MnP, Mito et al. developed a diagnostic approach to categorize magnetic domain dynamics into types 1--5, which account for both out-of-phase and nonlinear contributions to the ac magnetic response \cite{mit15}. Type 1 accounts for a purely linear dynamic response with no delay in phase against an ac field: $\gamma = 0, \eta =0$ in Eq.~\ref{eqn:Duffing}; type 2 dynamics introduces damping which leads to an opening of the ac hysteresis loop: $\gamma \ne 0, \eta =0$; for types $3-5$, $\gamma \ne 0, \eta \ne 0$, and the nonlinear term continuously grows. Remarkably type 5 dynamics display the smallest ac hysteresis, where \imm1(\imsuscept1{}$h$) $\sim0$. Thus, ac hysteresis loops categorized into types 1--5 notes the increasing dominance of the nonlinear contribution to the total magnetization dynamics \cite{mit15}, leading to \cpxm3 values in excess of 10\% of \cpxm1. The diagnostic approach was successfully applied to describe the unique domain dynamics in several other DMI-modulated systems \cite{mit09,mit12,mit15,tsu16b,tsu18}. 
	
	We now investigate in more detail the anisotropy-driven IC--C transitions by analyzing the nonlinear magnetic response to a time-dependent field via the temperature dependence of the third-order ac magnetization, \cpxm3. Figures \ref{fig:M3}(a) and \ref{fig:M3}(b) show \cpxm3{}$(T,H)$ for $f=1111$ Hz measured with an ac field amplitude $h = 5$ Oe and the associated surface plot. The dashed lines mark the phase boundaries determined from the linear susceptibility, as in Fig.~\ref{fig:loss}(b). For the SkL (Cyc), nonlinear response is absent in the pure phases, however, small values of \cpxm3 are restricted to the lossy regimes along the SkL$^{\mathsf{I/II}}$--Cyc$^{\mathsf{I/II}}$ and SkL$^{\mathsf{I/II}}$--FM (Cyc$^{\mathsf{I/II}}$--FM) phase boundaries. A similar trend was reported for MnSi, where the magnetization dynamics soften in a conical-SkL mixed-phase regime.\cite{tsu18} On the other hand, at $H = 0$ Oe, a massive anomaly in the nonlinear magnetic response (dashed arrow) extends from well within the Cyc phase from $T\approx8.75$ K down to $T<3$ K. The extended region reaches a maximum near $T_\mathrm{IC\rightarrow C}(0) =5.25$ K, similarly to \resuscept1 and \imsuscept1. The peak drops dramatically with applied magnetic field [\ref{fig:M3}(a)] and shifts slightly to lower temperatures, forming a dome of enhanced nonlinear response that expands into the FM regime of the phase diagram [\ref{fig:M3}(b)]. 

\begin{figure*}[]
\includegraphics[width=\linewidth]{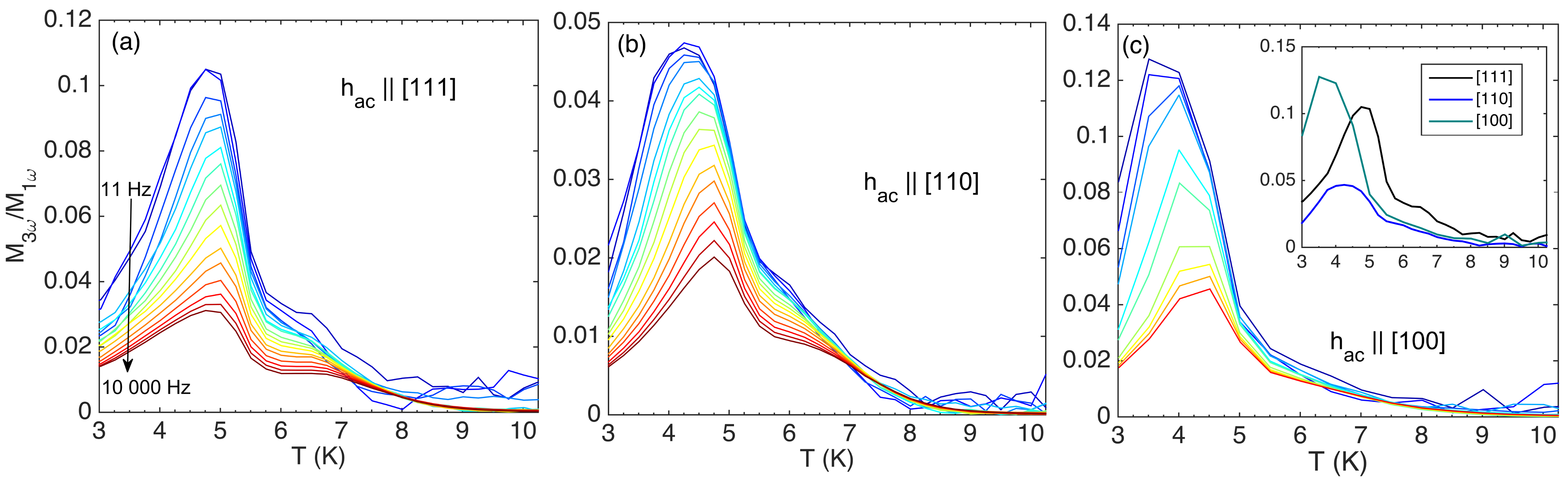}
\caption{\cpxm3/\cpxm1{}($T,0$) measured with $h_{ac} \| [111], [110], [100]$ for $f = 11 - 10000$ Hz.}
\label{fig:KlirrvT}
\end{figure*}

	Figure \ref{fig:KlirrvT}(a--c) plots the frequency dependence of \cpxm3{}($T$,0) scaled by the harmonic magnetic response, \cpxm1{}($T$,0), for ac field applied along each of the three principle cubic axes, [111], [110], and [100], measured on warming using a zero-field cooling protocol across the Cyc--FM transition. The ratio \cpxm3/\cpxm1, called the Klirr factor, characterizes the relative strength of the nonlinearity in the overall magnetic response \cite{mit15}. In \gvs, the maximum magnitude of \cpxm3/\cpxm1{}($T$,0) occurs at the lowest frequency measured, $f = 11$ Hz, on the order of characteristic frequency of magnetic domain dynamics. The Klirr factor systematically drops as the time-window of the measurement narrows, which indicates that the anharmonic response of the Cyc--FM crossover is related to the dynamics of the spin structure over large lengthscales. 
	
	The temperature dependence of the nonlinear response also displays anisotropic behavior. For comparison, each \cpxm3/\cpxm1{}($T$,0) curve for $f = 11$ Hz is plotted in the inset of Fig.~\ref{fig:KlirrvT}(c). The magnitude and profile of the Klirr factor apparently depends upon the orientation of $h_{ac}$ relative to the easy axes of the domains. At zero-field, the Cyc--FM transition occurs within each of the four structural domain variants, and is presumably driven by strong easy axis anisotropy along the direction of rhombohedral distortion. In the $[110]$ orientation, where the easy axes of two of the four domains are perpendicular to the ac field modulation, the Klirr factor takes on a value of $\sim5\%$. This value is approximately {\it half} of the maximum Klirr values for $H \| [111]$ (11\%), where the remaining three $\left<111\right>$ directions span $71^{\circ}$, and for $H \| [100]$ (13\%), where all four domains span $55^{\circ}$ with $H$. Such dependence of the strength of the nonlinearity on the relative orientation between the magnetic easy axes and $h_{ac}$ is consistent with experimental observations that the Cyc spin textures have a well-defined propagation direction, namely along $\left<110\right>$ within the $\{111\}$ planes \cite{whi18}. The ac field component perpendicular to the modulation direction of the spin structure, that is, along the magnetic easy axis, likely contributes to the large anharmonic response, supporting the scenario that the strong nonlinearity results from a well-defined domain structure as the cycloid period expands.

\begin{figure*}[]
\includegraphics[width=.9\linewidth]{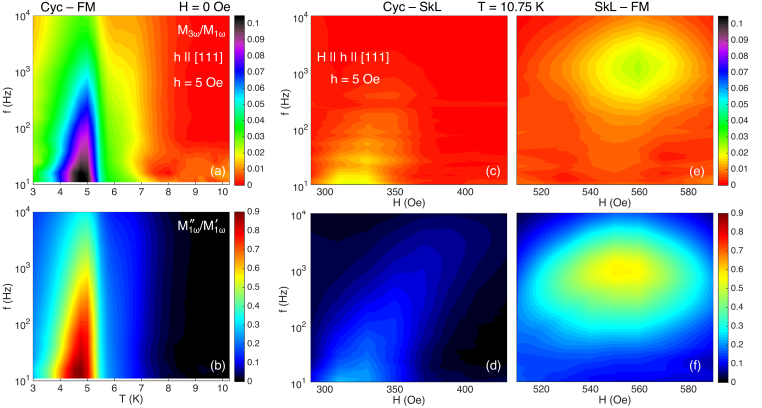}
\caption{Characterization of the magnetic domain dynamics using the Klirr factor, \cpxm3/\cpxm1, and loss tangent, $\tan\theta_{1\omega} =$ \imm1/\rem1, for several transition regimes. For $f = 11$ Hz, (a) \cpxm3/\cpxm1{}($T,f$) reaches 11\% and (b) $\tan\theta_{1\omega}(T,f)$ reaches 90\% across the extended zero-field Cyc--FM transition for $H \| [111]$. \cpxm3/\cpxm1{}($H,f$) and $\tan\theta_{1\omega}(H,f)$ for fields spanning the (c--d) Cyc$^\mathsf{I}$--SkL$^\mathsf{I}$ and (e--f) SkL$^\mathsf{I}$--FM mixed phase regimes across the phase boundaries at $T = 10.75$ K, as in Fig.~\ref{fig:Cole}.}
\label{fig:KlirrTanSurf}
\end{figure*}

	Figs.~\ref{fig:KlirrTanSurf}(a--b) present \cpxm3/\cpxm1{}($T,f$) and $\tan\theta_{1\omega}(T,f)$ across the zero-field Cyc crossover approaching the FM phase. The broad anomaly remains centered around $T_\mathrm{IC\rightarrow C}(0) = 5.25$ K up to the highest frequency and is maximized as the measurement window expands toward longer timescales. The enormous magnetic loss, of up to 90\% of the in-phase component at $f=11$ Hz, is also maximized at the longest timescale. The magnitude of $\tan\theta_{1\omega}$ lies far beyond the maximum predicted by the thermodynamic Cole-Cole model (50\%) \cite{bal03}. Figs.~\ref{fig:KlirrTanSurf}(c--f) display \cpxm3/\cpxm1{}($H,f$) and $\tan\theta_{1\omega}(H,f)$ for $T=10.75$ K, across the SkL boundaries that were analyzed in Sec. \ref{sec:dissmech} [Figs.~\ref{fig:Cole} and \ref{fig:skydyn}]. In (c--d), the Cyc--SkL transition exhibits a Klirr factor \cpxm3/\cpxm1{}(310 Oe, 11 Hz) $=2\%$ and $\tan \theta_{1\omega} = 0.24$. The loss tangent peak shifts toward higher frequency with increasing field and drops in magnitude. Here, the dynamics accelerate as the SkL dominates the mixed-phase regime. In (e--f), the Klirr factor for the SkL--FM transition is maximized for frequencies that coincide with the maximum $\tan\theta_{1\omega}$, reaching \cpxm3/\cpxm1{}(560 Oe, 1166 Hz) $=2.4\%$ with $\tan \theta_{1\omega} = 0.55$. \cpxm3{}($H,f$) peaks across the SkL--FM phase boundaries where glassy behavior is prominent, brought on by the softening of the SkL dynamics. The calculated values of \cpxm3/\cpxm1 and $\tan\theta_{1\omega}$ for various Cyc and SkL phase transitions are presented in Table~\ref{tab:table1} and are compared to other magnetic systems in the literature, as discussed in more detail in the following. 

\section{Discussion}
\subsection{Zero-field harmonic order}
In DMI magnets, the fixed rotation sense of the spin modulation results in robust textures that display long-range spin coherence. Additionally, the symmetry of the crystal lattice dictates the orientation of the DM vectors, thereby imparting anisotropy over the magnetic structure. The enhanced structural rigidity and long-range spin coherence can lead to an amplified nonlinear ac magnetic response. Table~\ref{tab:table1} reviews several relevant magnetic systems in the literature. For instance, \Cr exemplifies the CHM with a single DM vector; The spiral order with virtually infinite spin coherence propagates along the crystallographic $c$-axis and reaches \cpxm3/\cpxm1 $> 10 \%$ \cite{tsu16b,tsu18,cle18}. That the fixed rotation sense is essential to the magnitude of the nonlinear responses of different materials was first observed by Mito and coworkers in a study of a chiral molecule-based magnet, where the single-handed structure displayed a Klirr factor on the order of 10\% \cite{mit09}. Conversely, in the multidomain helimagnetic state of MnSi, where the propagation direction of the chiral helices changes in adjacent domains, a more recent study revealed a small nonlinear contribution ($\sim 0.2\%$) across the PM--CHM phase boundary \cite{tsu18}. In \gvs, the multidomain nature of the magnetism results from the structural domains brought about by the ferroelectric transition above \tc. Additionally, within domains it appears that the q-vectors of the spin cycloid may equally populate $\left<110\right>$ directions \cite{whi18}. These properties culminate into \cpxm3 = 0 at the PM--Cyc phase boundary in \gvs, as shown in Fig.~\ref{fig:M3}. However, as discussed in more detail later, despite the multidomain state, the anisotropy-driven Cyc--FM transition centered at $T_\mathrm{IC\rightarrow C}(0)$ displays an enormous nonlinear response.

\subsection{Skyrmion lattice}
The solitonic structures that emerge out of the harmonically modulated ground (or zero-field) states also display robust spin coherence, such as the 1D chiral soliton lattice or the 2D skyrmion lattice, and have distinct dynamic signatures. The chiral soliton lattice (CSL), a periodic system of uniaxially propagating ferromagnetic domains separated by $2\pi$ domain walls, also displays enormous \cpxm3 \cite{tsu16b,tsu18,cle18}, but is accompanied by magnetic loss due absorption of energy as the domain walls overcome energy barriers. However, as shown in the case of MnSi, \cpxm3 is absent within the highly coherent 2D skyrmion lattice phase pocket \cite{tsu18}. Here the timescale of the pure phase spin dynamics is much faster than the timescales related to typical domain dynamics; As shown in ac susceptibility studies of CHMs and polar \gvs, the susceptibility approaches the adiabatic limit ($\tau \rightarrow 0, f \rightarrow \infty$) on entering the single-phase SkL and Cyc regimes \cite{lev14,but17a}. Notably, in MnSi, a modest \cpxm3/\cpxm1 ($\sim1\%$), approximately one order of magnitude larger than the CHM-conical transition, emerges only along the phase boundaries where the skyrmion domains soften and magnetic dissipation appears \cite{tsu18}. As shown in the surface plot in Fig.~\ref{fig:M3}(b), \cpxm3 appears at the Cyc--SkL ($\sim2\%$ of \cpxm1 in Fig.~\ref{fig:KlirrTanSurf}) and SkL--FM ($\sim2.4\%$ of \cpxm1) boundaries where glassy dynamics were observed in Fig.~\ref{fig:Cole}. 

Furthermore, the spatial rigidity of the SkL was found to be much lower than that of the CSL in Ref. \cite{tsu18}, yielding a Klirr factor on the order of $10^{-2} - 10^{-3}$, similar to spin and cluster glass states. In MnSi, the direction of the applied magnetic field dictates the propagation direction of the q-vector, leading to a reorientation into a single helical domain across the CHM to conical magnetic crossover boundary \cite{bau17}. Depending on the magnetic field direction, the skyrmion cores in cubic CHMs may align along, e.g. $\left<111\right>$, $\left<110\right>$, $\left<100\right>$, in a single domain structure. In \gvs, skyrmion cores are confined along the easy axis of magnetization \cite{leo17}, which likely provides stronger spatial rigidity in comparison to the B20 family . However, in the multidomain configuration, the apparent nonlinear response is on the order of $10^{-2}$. Achieving a single domain SkL state in \gvs could drive the Klirr ratio toward higher values than in SkL hosts with weak anisotropy. 

\subsection{Cycloid crossover}

\begin{table}
 \caption{ \label{tab:table1}Maximum Klirr factor, \cpxm3/\cpxm1, and loss tangent, $\tan \theta_{1\omega} = $ \imm1/\rem1, values observed in \gvs compared to select systems exhibiting robust spin textures stabilized via the Dzyaloshinskii-Moriya interaction.}
 	\begin{ruledtabular}
 		\begin{tabular}{lcccc}
			Regime&$M_{3\omega}/M_{1\omega}$&$M^{\prime\prime}_{1\omega}/M^{\prime}_{1\omega}$&Type\footnote{based on the prescription developed in \cite{mit15}}&Source\\
			\hline 
			& \\[\dimexpr-\normalbaselineskip+2pt]
			\gvs\\
			Cyc$\rightarrow$FM    & $0.13$                   & $0.97$    & 4\\
			Cyc$\rightarrow$SkL    & $0.02$                   & $0.24$    & 3& this work\\
			SkL$\rightarrow$FM    & $0.024$                  & $0.55$    & 3\\
			PM$\rightarrow$Cyc    & $0$                        & $0$         & 1 \\
			& \\[\dimexpr-\normalbaselineskip+1pt]
			\hline
			& \\[\dimexpr-\normalbaselineskip+2pt]
			MnSi\\
			SkL$\rightarrow$Conical      & $0.01$                    & $0.06$   & 3& \cite{tsu18}\\
			PM$\rightarrow$CHM          & $0$                          & $0$       & 1 \\
			& \\[\dimexpr-\normalbaselineskip+1pt]
			\hline
			& \\[\dimexpr-\normalbaselineskip+2pt]
			\Cr\\
			PM$\rightarrow$CHM                & $0.1$                       & $0$         & 5 & \cite{tsu16b}\\
			HNL CSL\footnote{highly nonlinear chiral soliton lattice}               & $0.13$                    & $0.5$        & 4 & \cite{tsu16b,cle18}\\
			& \\[\dimexpr-\normalbaselineskip+1pt]
			\hline
			& \\[\dimexpr-\normalbaselineskip+2pt]
			MnP\\
			PM$\rightarrow$FM            & $0.1$                       & $\sim0$          & 5 & \cite{mit15}\\
			IT striped domain\footnote{intermediate temperature (IT) structure and DM interaction identified in \cite{yam14}} & $\sim0.2$                       & $\sim0.8$       & 4\\
			& \\[\dimexpr-\normalbaselineskip+1pt]
			\hline
			& \\[\dimexpr-\normalbaselineskip+2pt]
			R-GN\footnote{chiral molecule-based magnet $\mathrm{[Cr(CN)_6][Mn}(R)$-$\mathrm{pnH(H_2O)](H_2O)}$} \\
			PM$\rightarrow$FiM          & $\sim0.16$              & $\sim0$           & 5 & \cite{mit12}\\
		\end{tabular}
	\end{ruledtabular}
\end{table}

At $H = 0$ Oe, the anisotropy-driven incommensurate Cyc to commensurate FM transition displays anomalous ac susceptibility, magnetic loss, and nonlinear dynamic response to the time-dependent magnetic field in the range 9 K $> T >$ 3 K centered around the peak at $T_\mathrm{IC\rightarrow C}(0) = 5.25$ K. At $T = 7$ K, \cpxm3/\cpxm1 is nearly $2\%$ for $H \| [111]$ which corresponds to the change in slope of the temperature dependence of the cycloidal scattering vector in \cite{whi18}.  There, the scattering vector decreases more rapidly and falls outside the detectable limit as temperature is lowered to $T = 4.5$ K. Thus, our results suggest that a precipitous growth in the Klirr factor accompanies a more rapid increase in the cycloid wavelength. Presumably, the spatially modulated structure becomes more nonlinear as the cycloid pitch diverges toward the IC--C transition, as pointed out by White and coworkers in \cite{whi18}. Such a process was described theoretically to occur via a soliton lattice according to Izyumov \cite{izy84}. Interestingly, line-shape analysis of $^{57}$Fe nuclear magnetic resonance spectra revealed the solitonic character of the the cycloid spin profile in the quintessential multiferroic $\mathrm{BiFeO_3}$ \cite{zal00}. In $\mathrm{BiFeO_3}$, the cycloid becomes more anharmonic on lowering temperature \cite{zal00} and, based on electron spin resonance results in \cite{rue04}, increasing magnetic field. In the present study on \gvs, the extended range of the nonlinear behavior and strong magnetic loss suggests that the zero-field IC--C transition occurs via a broad crossover process involving a correlated domain arrangement. However, the exact form of the magnetic structure remains to be determined.

The maximized Klirr factor of up to $13\%$ for $H \| [100]$ reaches a value on the order of the highest reported in DMI magnets, as shown in Table~\ref{tab:table1}. In a previous study, we observed Klirr factors over 10\% for the highly nonlinear chiral soliton lattice state in \Cr \cite{cle18}. There the susceptibility peak does not correspond to the metamagnetic phase transition, but it represents the center of an extended nonlinear crossover process that terminates with an IC--C phase transition when \cpxm{(n>1)} falls to zero.\cite{cle18} The loss tangent, $\tan\theta_{1\omega}$ = \imm1/\rem1, at $T_\mathrm{IC\rightarrow C}(0)$ is also enormous, ranging from $86 - 97\%$.  Based on the prescription developed by Mito, et al. \cite{mit15} to categorize ac responses of unique magnetic domains, the dynamics fall into the type 4 class categorized by \cpxm3/\cpxm1 $\ge$ 0.05 and $\tan \theta_{1\omega} \ne 0$. Studies of \Cr by Tsuruta et, al. place the highly nonlinear chiral soliton lattice to FM transition, MnP, and R-GN at type 4, whereas the single domain PM--CHM transition dynamics behave as type 5 with no \imm1 \cite{mit09,mit12,mit15,tsu16b}. As a large Klirr factor for frequencies at or below $\sim10$ Hz is typically associated with the dynamic character of magnetic domain formation, the results herein support the notion that a robust and unique domain structure could emerge out of the Cyc state and the topic deserves further study via theoretical modeling and neutron and microscopy techniques.

In Ref.~\cite{tsu18}, Tsuruta et al. note that the large anharmonicity in the magnetic response due to coherent domain formation is typically observed to peak at the boundaries of phase transitions, where it is attributed to thermal softening of spin dynamics. Anharmonicity is not observed at the PM--Cyc phase boundary in \gvs despite SANS evidence that the wave vector continuously decreases from 13 K in zero field.  Such behavior is not surprising as multiple DM vectors govern the modulated state, leading to cycloids distributed over $\left<110\right>$, as well as the inherent multidomain nature that breaks the spin coherence at structural domain boundaries. In the B20 helimagnets, cubic anisotropy leading to the multidomain zero-field CHM phase, and the resulting presence of multiple DM axes, was attributed to the weak nonlinear response at the CHM--PM and CHM--conical transitions \cite{tsu18}. However, in stark contrast, the Cyc--FM crossover displays enormous nonlinear response as the magnetocrystalline anisotropy becomes the dominant energy scale at low temperatures, thereby enhancing the rigidity of the spin structure despite its multidomain nature. 

\section{Conclusion}

The dissipation mechanisms and nonlinear magnetization dynamics were analyzed across the magnetic phase diagram of \gvs via \suscept1{}$(T,H)$ and \cpxm3{}$(T,H)$ for $f = 10-10000$ Hz. Dynamic signatures of mixed-phase behavior were observed across the Cyc--SkL and SkL--FM transitions, consistent with dc magnetic hysteresis observed between warming and cooling curves. The ratio \cpxm3/\cpxm1characterizing the nonlinear response was calculated for the strongly pinned \N SkL. Similar to results in MnSi, nonzero \cpxm3 only occurred at the phase boundaries where SkL dynamics soften. However, even with a multidomain configured \N SkL, \cpxm3/\cpxm1 values (Cyc--SkL: 2\%, SkL--FM: 2.4\%) were slightly enhanced with respect to recent results in MnSi (conical--SkL: \%1), reflecting the influence of anisotropy on the rigidity of the spin texture. 

The frequency dependence of Cyc--FM transitions for $H >200$ Oe and $H<200$ Oe were compared to reveal separate relaxation mechanisms. \cpxm3{}$(T,H=0)$ illustrates that the IC Cyc displays enormous anharmonicity in its dynamic response when the magnetocrystalline anisotropy becomes the dominant energy scale at low temperatures, thereby increasing the rigidity of the spin structure. \cpxm3/\cpxm1 depends strongly on the ac field orientation and for $H\|[100]$ reaches 13\% at $T_\mathrm{IC\rightarrow C}(0) = 5.25$ K, previously identified in the literature as the FM critical field. Combined with the observed anomalous magnetic loss tangent $\sim 90\%$, our results support the notion that a robust and unique domain structure emerges out of the Cyc state across the IC--C transition.

\begin{acknowledgments}
Research at the University of South Florida was supported from the U.S. Department of Energy, Office of Basic Energy Sciences, Division of Materials Sciences and Engineering under Award No. DE-FG02-07ER46438. M.H.P. also acknowledges support from the VISCOSTONE USA under Award No. 1253113200. G. P. and D. M. acknowledge support from the National Science Foundation under grant DMR-1410428. A. C. was supported by the U.S. Department of Energy (DOE), Office of Science, Basic Energy Sciences, Materials Sciences and Engineering Division. Research at the Naval Research Laboratory was funded by the Office of Naval Research (ONR) through the Naval Research Laboratory Basic Research Program.
\end{acknowledgments}


%

\end{document}